\documentclass{edm_article}
\usepackage{float}
\usepackage{svg}
\usepackage{amsmath}
\usepackage{comment}
\usepackage{hyperref}
\usepackage{multirow}
\usepackage{makecell}
\usepackage{booktabs}
\usepackage{balance}
\begin{document}

\title{Generalisable Methods for Early Prediction in \\ Interactive Simulations for Education}


\newcommand{\jade}[1]{\textcolor{black}{#1}}

\author{
Jade Maï Cock\\
      \affaddr{EPFL}\\
      \email{\normalsize{jade.cock@epfl.ch}}
      \and
Mirko Marras\\
      \affaddr{University of Cagliari}\\
      \email{\normalsize{mirko.marras@acm.org}}
      \and
Christian Giang\\
      \affaddr{EPFL}\\
      \email{\normalsize{christian.giang@epfl.ch}}
      \and
Tanja Käser\\
      \affaddr{EPFL}\\
      \email{\normalsize{tanja.kaeser@epfl.ch}}
}

\maketitle

\begin{abstract}
Interactive simulations allow students to discover the underlying principles of a scientific phenomenon through their own exploration. Unfortunately, students often struggle to learn effectively in these environments. Classifying students' interaction data in the simulations based on their expected performance has the potential to enable adaptive guidance and consequently improve students' learning. 
Previous research in this field has mainly focused on a-posteriori analyses or investigations limited to one specific predictive model and simulation.
In this paper, we investigate the quality and generalisability of models for an early prediction of conceptual understanding based on clickstream data of students across interactive simulations. We first measure the students' conceptual understanding through their in-task performance. Then, we suggest a novel type of features that, starting from clickstream data, encodes both the state of the simulation and the action performed by the student. We finally propose to feed these features into GRU-based models, with and without attention, for prediction. Experiments on two different simulations and with two different populations show that our proposed models outperform shallow learning baselines and better generalise to different learning environments and populations. The inclusion of attention into the model increases interpretability in terms of effective inquiry. The source code is available on Github\footnote{https://github.com/epfl-ml4ed/beerslaw-lab.git}.
\end{abstract}


\section{Introduction}
In the past years, interactive simulations have gained increasing popularity in formal education \cite{bo2018secondary} and have become integral parts of many science curricula \cite{rutten2012learning}. These environments provide visualisations of abstract concepts, which can help students to better grasp them \cite{wu2007ninth,mcelhaney2011investigations}. On the other hand, these simulations also allow students to explore new learning content and build knowledge in an active and independent way \cite{wieman2008phet}. Since simulations are virtual learning environments, students can experiment freely without fearing any serious consequences of making mistakes - this property can often make them a preferred choice over real lab sessions. It has also been proved that the use of interactive simulations can support inquiry-based learning \cite{moore2013interactive}. 

However, previous work has pointed out that many students struggle to learn effectively with interactive simulations \cite{alfieri2011does,kirschner2006unguided,mayer2004should,cock2021early}. Indeed, navigating through interactive simulations purposefully can be challenging, especially when the number of controls and the level of complexity are high \cite{adams2008levels}. Therefore, providing adaptive guidance and feedback has the potential to improve students' learning outcomes. However, interactive simulations are often highly complex systems with a practically unlimited number of possible learning paths. Hence, defining (un-)productive inquiry behaviour in such environments is still an open issue.

Prior work has focused on identifying the key factors of productive inquiry behaviour in interactive systems from student log data by leveraging sequence mining and clustering techniques. For instance, \cite{perez2017identifying} applied information theoretic sequence mining to identify behavioural differences of students using a virtual simulation for electronic circuits. In \cite{baker2016towards}, the authors trained binary classifiers and logistic regression models on log data from virtual environments to categorise students by their science inquiry skills. Similarly, \cite{perez2018control} used student log data as input for a linear regression model to predict the conceptual understanding students acquired after using physics and chemistry simulations. Latent Class Analyses models were able to identify different profiles in inquiry performance in two PISA science assessments involving interactive simulations \cite{teig2020identifying}.
The work in \cite{kardan14usefulness} employed clustering techniques to analyse students' inquiry strategies in an interactive simulation on circuit construction. Log data was manually tagged as ground truth to train a classifier on successive inquiry behaviour in \cite{gobert13}. Other research focused on the teacher and provided them with a dashboard displaying the mined student sequences \cite{tavares2019towards}. A rule-based approach for classifying students' problem solving strategies in an interactive Chemistry simulation is described in \cite{gal15}. In recent work, \cite{wang2021automating} assessed the effectiveness of data mining techniques guided by knowledge gained from qualitative observations to predict student's problem-solving skills in an interactive simulation. 

However, most of the existing approaches have performed \textit{a posteriori} analyses, requiring the complete log data entries as input to the models. This represents a major limitation, since it does not allow the system to provide feedback to the students while they are still interacting with the simulation. Their high complexity and lack of predefined learning trajectories makes building a student model for these environments a challenge. A popular approach to adaptive interventions in open ended learning environments is the use of a clustering-classification framework \cite{kardan2011framework}. Students are clustered offline based on their interaction sequences and the resulting clusters are interpreted. The classification step is then done online: students are assigned to a cluster in real-time, as explored in \cite{fratamico2017applying,conati15representations} for an interactive simulation for electrical circuit construction. Recently, \cite{cock2021early} presented a novel pipeline for the early prediction of conceptual understanding in an interactive Physics simulation. They were able to robustly predict students' conceptual understanding in an interactive electronics simulation with only initial fractions of the log data. However, the approach was evaluated using only one data set from a single population (Physics undergraduates). \jade{To make such approaches more impactful, it becomes crucial to enable their generalisation across simulations and environments. }


In this paper, we propose a new generalisable approach for predicting students' conceptual understanding based on sequential log data from interactive simulations. We leverage students' interaction data to extract meaningful features that encode both the state of the simulation and the action of the student. These features are then used as an input to models based on Gated Recurrent Units (GRUs), to predict students' conceptual understanding. In addition to a standard GRU model, we also propose an extension, the Self-Attention-based Gated Recurrent Unit model. We extensively evaluate our approach on two data sets from different contexts: the first data set stems from students of $10$ different vocational schools interacting with a chemistry simulation. The second data set consists of physics undergraduate students using a physics simulation. With our experiments, we aim to address three research questions: 1) To what degree is interaction data within interactive simulations predictive for their obtained conceptual understanding? 2) Can our models predict students' conceptual understanding early on? 3) Is our approach transferable to another context involving a different population and simulation?

Our results demonstrate that our proposed models outperform previous approaches in terms of (early) predictive performance on both data sets. We further show that the inclusion of self-attention enables us to draw conclusions about productive inquiry behaviour in both simulations.

\section{Learning Environments}\label{section:contextdata}
The PhET interactive simulations\footnote{https://phet.colorado.edu/} allow students to explore natural phenomena using different parameter configurations and ideally infer the underlying principles on their own. A large body of research has focused on analysing students' inquiry strategies using these simulations, usually coupled with an explicit task to be solved using the simulation \cite{yehya2019learning}. To evaluate the (early) prediction models presented in this paper, we used a ranking task which we adapted to two different simulation environments. \jade{For each ranking task, the values of the variables presented in the task are outside the range provided by the simulation. Hence, students need to inquire the system to uncover the appropriate equations and plug in the numbers as answers in the ranking task.}

\begin{figure}[!t]
\centering 
\includegraphics[width=0.9\columnwidth]{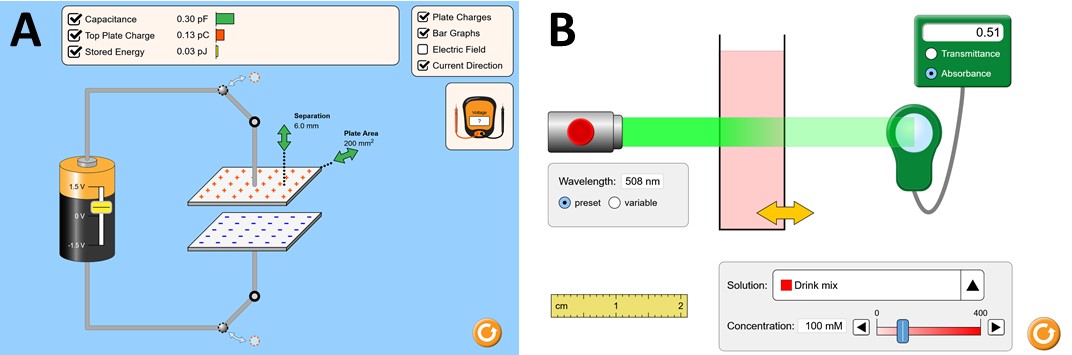} 
\caption{The two learning environments used in this study, the Capacitor Lab (A) and the Beer's Law Lab (B).} 
\label{fig:phet-envs} 
\end{figure}

\begin{figure}[!t]
\centering
{\includegraphics[width=\columnwidth]{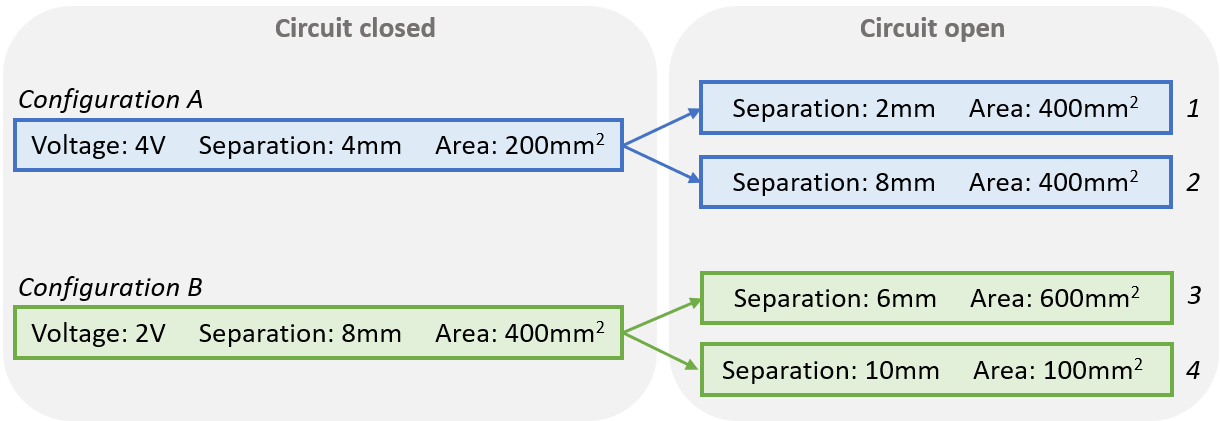}}\\
{\includegraphics[width=\columnwidth]{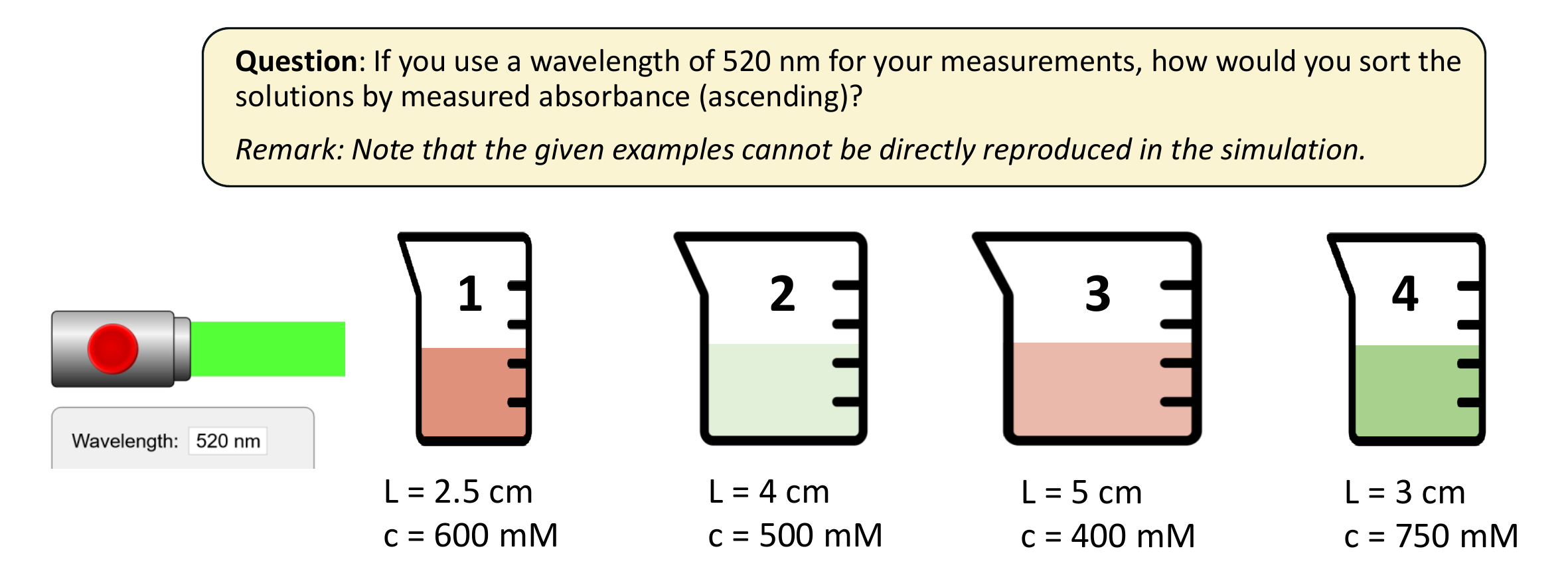}}\label{fig:ranktaskchemlab}
\caption{Ranking tasks for the two interactive simulations.}
\label{fig:ranktask}
\end{figure}

\textbf{Capacitor Lab}. The PhET Capacitor Lab simulation\footnote{https://phet.colorado.edu/en/simulations/capacitor-lab} provides students with the possibility to manipulate different parameters of an electric circuit (plate area, plate separation, applied voltage, open/close the circuit), to help them understand how each parameter influences the capacitance of a plate capacitor as well as the energy stored in that capacitor (see Fig. \ref{fig:phet-envs}A). Previous work \cite{cock2021early} designed an inquiry-based learning activity for this simulation using a ranking task. Specifically, students were asked to rank four different configurations according to the stored energy saved in the plate capacitor (see Fig. \ref{fig:ranktask}a). The four configurations were created by first loading the plate capacitor using two parameter configurations in a closed circuit (A and B), followed by opening the switch to disconnect the capacitor from the circuit and changing the parameters to obtain the final configurations to be ranked (1, 2, 3, and 4). The authors also suggested a tree-based approach to label the $24$ possible solutions based on conceptual understanding. For this paper, we will use the binary labels suggested by \cite{cock2021early} to divide students into a group with an \texttt{advanced understanding} of stored energy (i.e., for both open and closed circuits) and a second group with a conceptual understanding limited to the closed circuit, denoted as \texttt{limited understanding}.

\begin{figure}[t]
\centering
\includegraphics[width=\columnwidth]{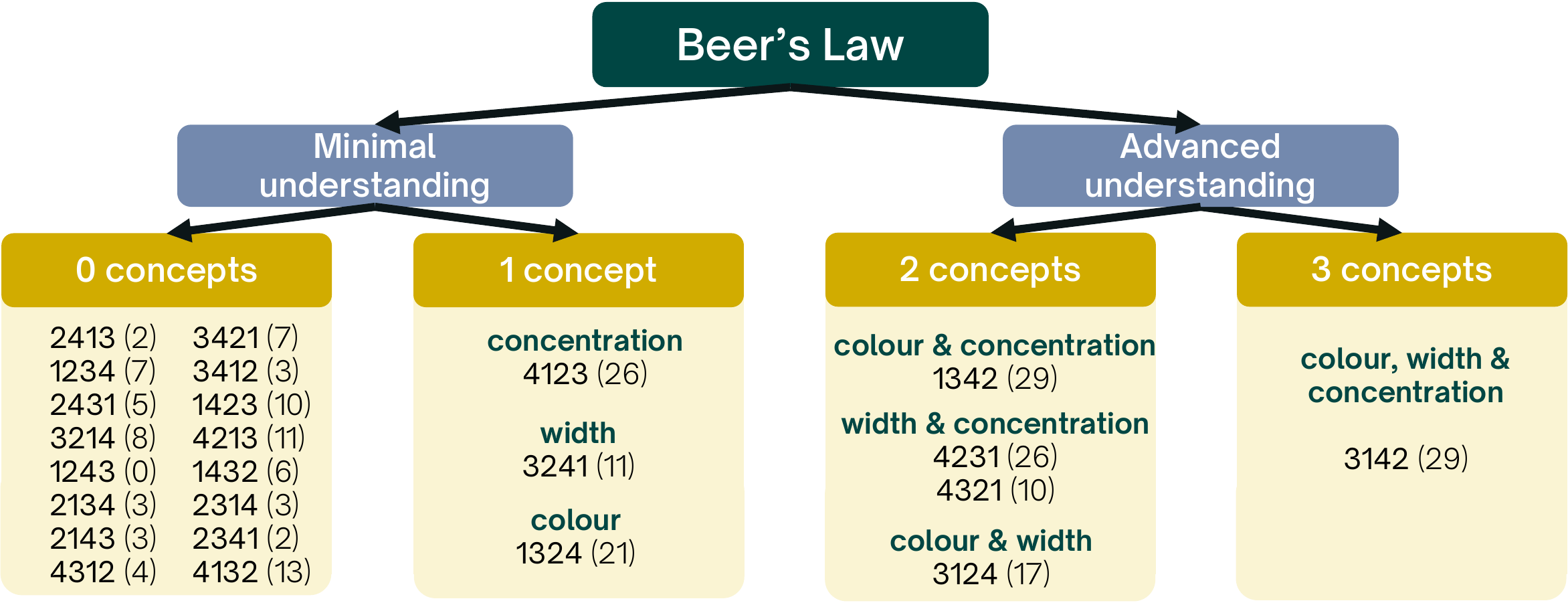}
\caption{The tree used to map the rankings to different levels of conceptual understanding. The numbers in parentheses indicate the number of submissions for the ranking.}
\label{fig:labelschemlab}
\end{figure}

\textbf{Beer's Law Lab}. The PhET Beer's Law Lab simulation\footnote{https://phet.colorado.edu/en/simulations/beers-law-lab} allows students to explore how light is absorbed in a solution and which factors influence the measured absorbance (Fig. \ref{fig:phet-envs}B). In this environment, the students can manipulate the following components: the wavelength $\lambda$ of the light beam, the width of the container (representing the light's path length $L$ through the solution), the substance of the solution (changing its colour), and its concentration $c$. By experimenting with these components, in the ideal case, students can infer Beer's Law, which describes how each parameter influences the measured absorbance: $ A = \epsilon_\lambda * L * c $. Based on the inquiry-based learning task presented by \cite{cock2021early}, we developed a ranking task that can be solved with Beer's Law Lab. Students were presented with four flasks that varied in width as well as in the color and concentration of the solution they contained (Fig. \ref{fig:ranktask}b). Given a light beam of a specific wavelength, students were asked to rank the four flasks by the measured absorbance. It is important to note that the values chosen for the four flasks were not available in the simulation in order to prevent students from obtaining the correct ranking by simply replicating the flasks. To label the submitted rankings, we used the tree-based approach introduced in previous work \cite{cock2021early}. We classified each of the $24$ possible rankings according to the level of conceptual understanding (Fig. \ref{fig:labelschemlab}). Specifically, for each ranking, it was determined whether the influence of i) the substance's concentration, ii) the container's width and iii) the colour difference between light beam and solution was understood or not. Rankings were then grouped by the number of concepts understood that they represented. Rankings representing $0$ or $1$ concepts understood were then labelled as \texttt{limited understanding}, while rankings representing $2$ or $3$ concepts were labelled as \texttt{advanced understanding}.

\section{Method}
\label{sec:method}
\begin{figure*}[t]
\centering
\includegraphics[trim= 0.0 0.0 0cm 4.5cm,width=0.8\textwidth]{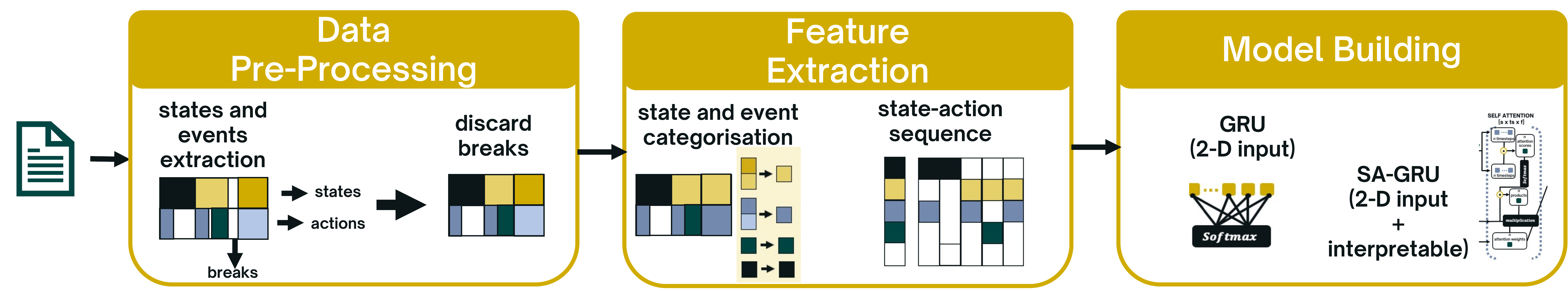}
\caption{Our model building pipeline consists of three steps: we first extract the logs into sequences of events and states. We then categorise the states and events to build state-action sequences. Finally we feed the features into variations of GRU models.}
\label{fig:datapipeline}
\end{figure*}

We are interested in creating \textit{generalisable models} to (early) predict the level of conceptual understanding students will achieve after interacting with a simulation. Formally, our goal is to predict, for each student, a binary label representing their level of conceptual understanding. To address this binary classification task, we propose a pipeline consisting of the four steps illustrated in Fig. \ref{fig:datapipeline}. We first pre-process the data into event logs and extract so-called state-action sequences from these event logs. We then build different types of deep neural network models to perform the classification task. In what follows, we first formalise the addressed problem, before describing each step of the pipeline in detail.

\begin{figure*}[!t]
\centering
\includegraphics[trim= 0.0 0.0 0cm 4.5cm, clip,width=0.7\textwidth]{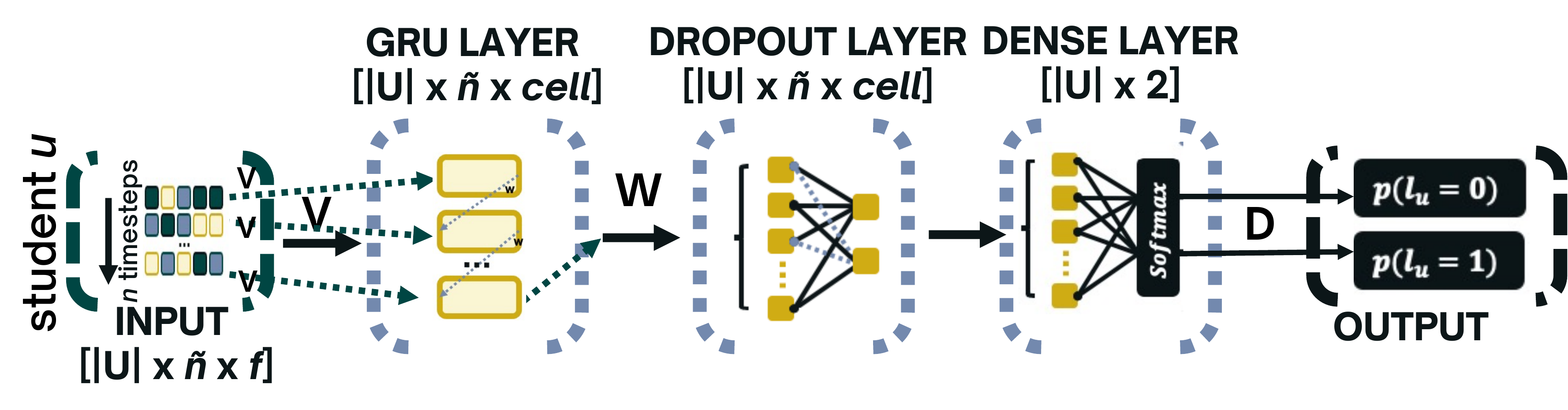}
\caption{Architecture of our \texttt{GRU} model with one unrolled GRU layer: the input features are fed sequentially into the recurrent cells of the GRU layer. The output shape of each layer is given: $|U|$ denotes the number of users, $\tilde{n} = \max_{u\in\mathcal{U}} \tilde{n}_u$, $cell$ the number of GRU cells, and $f$ the number of features in the input matrix.}
\label{fig:grudl}
\end{figure*}

\subsection{Problem Formalisation}
We consider $\mathcal{U}$ to be the set of students participating in the learning activity. During this learning activity, students use an interactive simulation in order to rank four different configurations according to a specific criterion (see Section \ref{section:contextdata} for details on the ranking tasks). We denote the set of possible answers to the ranking task as $\mathcal{A}$ and the answer of a student $u \in \mathcal{U}$ to this ranking task $a_u \in \mathcal{A}$. Based on their answers, we assign a label $L_u \in \{0, 1\}$ to each student $u \in \mathcal{U}$. This mapping $m$ is a direct function of students' answers and can be formalised as  $m: a_u\in \mathcal{A} \longmapsto L_u \in \{0, 1\}$ for $u\in \mathcal{U}$. For both lab activities, label $1$ denotes \texttt{advanced understanding} and label $0$ \texttt{limited understanding}.

\par To solve the ranking task, students interact with the simulation and investigate the factors influencing the dependent variable (absorbance in case of Beer's Law Lab, stored energy in case of Capacitor Lab). Their full and non-processed sequence of events in the simulation is denoted as:
\vspace{-5mm}

\begin{equation}\label{eq:rawsequence}
    R_u = \{(e_0, s_0, d_0), ..., (e_{n_u-1}, s_{n_u-1}, d_{n_u-1})\}
\vspace{-2mm}
\end{equation}

where $e_m$ is the $m^{th}$ \textit{event} $e$ , $s_m$ is the \textit{state of the simulation} at time step $m$, and $d_m$ denotes the \textit{duration of event} $e_m$ in \textit{state} $s_m$.
We denote the partial sequence of student $u \in\mathcal{U}$ from the first to the $l^{th}$ event as $R_u^l=\{(e_0, s_0, d_0), ..., (e_{l-1}, s_{l-1}, d_{l-1})\}$ with the value $l\in[1, n_u)$.
To clarify the notion of \textit{duration} and \textit{time steps}, we introduce $\|R_u\|_{duration} = t_u$ where $t_u$ denotes the total time in seconds spent by student $u$ on the simulation. We then denote $\|R_u\|_{timestep} = n_u$ where $n_u$ denotes the total number of events performed by student $u$ in sequence $R_u$.

Hence, training an (early) students' conceptual understanding predictor $\mathcal{E}$ with interactions $R_u$ and conceptual label $L_u$ after a number of interactions $\bf{n}_u$ becomes an optimisation problem, aimed to find model parameters $\theta$ that maximise the expectation on the following objective function (i.e., predict the correct label, given interactions) over a dataset $\mathbb G$:
\vspace{-10mm}

\begin{equation}
\tilde{\theta} = \underset{\theta}{\operatorname{argmax}} \mathop{\mathbb{E}}_{(R_u, L_u)\in \mathbb{G}}  L_u = \mathcal{E}(R_u^{\bf{n}_u} \;|\; \theta)
\label{eq:problem-definition}
\vspace{-2mm}
\end{equation}

In what follows, we describe the steps carried out to solve this problem from pre-processing until model building.  

\subsection{Data Pre-Processing}
From the log data, we extract the sequence of events of a student $u$ as follows: anything between a mouse click and a mouse release qualifies as an \textit{action}. Each \textit{action} can be described by its \textit{duration} and the \textit{component} the action is performed on, and is conducted in a specific \textit{simulation state}. Anything between a mouse release and a mouse click qualifies as a \textit{break}. Similarly, \textit{breaks} are conducted in a specific \textit{state} and have a specific \textit{duration}.
All \textit{actions} and \textit{breaks} generated by a student $u\in\mathcal{U}$ are sorted chronologically to form the raw interaction sequence of a user $u\in\mathcal{U}$, $R_u$ as per Eq. \ref{eq:rawsequence}, where events can be both \textit{actions} or \textit{breaks}.

Some \textit{breaks} in student interactions can be indicative of reflective, observational or distracted periods. Other \textit{breaks} can occur naturally throughout a student's interaction when the mouse moves from one element to the other \cite{wang2021examining}. The latter might introduce undesired noise and hinder student classification over their conceptual understanding, not unlike stop words negatively influence sentiment analysis tasks in the domain of natural language processing, as an example \cite{ghag2015comparative}. To maximise the retention of informative periods of inactivity, we therefore discard for each student $u\in\mathcal{U}$ their $60\%$ of shortest breaks. We are left with $R_u = \{(e_0, s_0, d_0), ..., (e_{\Tilde{n}_u-1}, s_{\Tilde{n}_u-1}, d_{\Tilde{n}_u-1})\}$ with $\Tilde{n}_u \leq n_u$, the final sequence of triplets of events $e_m$ occurring in state $s_m$ and of duration $d_m$ for student $u\in\mathcal{U}$ at time $m\in[0, t_u)$. \jade{The new sequence contains $\tilde{n}_u$} event triplets, i.e., \textit{timesteps}.

\begin{figure*}[!t]
\centering
\includegraphics[width=0.86\textwidth]{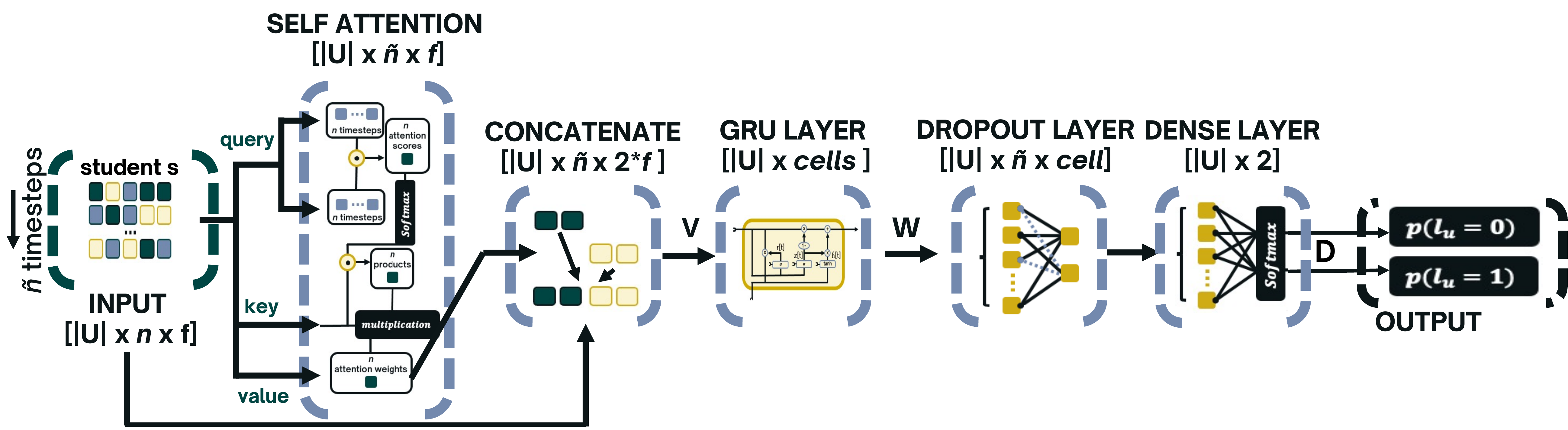}
\caption{\texttt{SA-GRU}'s architecture with the attention mechanism. The output shape of each layer is given: $|U|$ denotes the number of students, $\tilde{n} = \max_{u\in\mathcal{U}} \tilde{n}_u$, $f$ is the number of dimension in the state-action features, and $cells$ is the number of GRU cells.}
\label{fig:sagrudl}
\end{figure*}

\subsection{Feature Extraction}
\label{sec:featextract}
\jade{The data logging of the PhET environments captures any click targeting an "interactive" component as well as the state of the simulation in case one of these "clicks" changes the value of a variable/multiple variables in the system. This fine-grained logging results in high dimensional simulation states $s$ and a high number of different events $e$. Before creating feature vectors for each time step $m$, we therefore categorise the obtained events and states into meaningful components. We divide all possible \textit{simulation states} into two groups: a) helpful or b) unhelpful to solve the problem. We then ponder whether some of the \textit{helpful states} can be grouped such that the level of information received by the student is the same independent of the state within the group. We proceed in a similar way for the events. While this categorisation procedure is generalisable over different platforms, whether a simulation state or event is helpful depends on the actual task.}

\jade{\textbf{State-Action Categorisation for Capacitor Lab}. For the Capacitor Lab activity, we use the state and action categorisation defined in \cite{cock2021early}. By applying the above pipeline, we qualify each state $s_m \in (e_m, s_m, d_m)$ into one of the following groups: 1) closed circuit - stored energy displayed, 2) open circuit - stored energy displayed, 3) closed circuit - stored energy not displayed, and 4) open circuit - stored energy not displayed. Since stored energy is the dependent variable in the ranking task, displaying the stored energy is essential for solving the task. The relationship between the parameters (plate area, plate separation, battery voltage) and stored energy differs for closed and open circuits. Moreover, we map each event $e_m \in (e_m, s_m, d_m)$ into one of the following actions: 1) manipulating battery voltage, 2) manipulating plate area, 3) manipulating plate separation, 4) break, 5) other (any other action). }


\vspace{-1mm}
\jade{\textbf{State-Action Categorisation for Beer's Law Lab}. There are four independent variables that influence the dependent variable \textit{absorbance} of the ranking task: the \textit{laser colour}, the \textit{solution colour}, the \textit{concentration} of the solution, and finally the \textit{width} of the container. Consequently, we categorise each state $s_m \in (e_m, s_m, d_m)$ into one of those 4 groups: 1) green-green (absorbance: displayed, laser colour: green, solution colour: green), 2) green-red (absorbance: displayed, laser colour: green, solution colour: red), 3) absorbance displayed (absorbance: displayed, laser colour: not green and/or solution colour: neither red nor green), 4) not observed (absorbance: not displayed). To solve the ranking task depicted in Fig. \ref{fig:ranktaskchemlab}, students need to understand how the \textit{width} and \textit{concentration} interact with the \textit{absorbance} when the \textit{solution colour} is either red or green and when the \textit{laser colour} is green, i.e., states 1) and 2). The \textit{width} and the \textit{concentration} share a linear dependency with the \textit{absorbance} and are both colour agnostic. As long as the absorbance is observed, those relations can be discovered no matter the colour of the laser or the solution, i.e., state 3. Furthermore, we process each event $e_m \in (e_m, s_m, d_m)$ into one of those categories: 1) width (width slider is moved), 2) concentration (concentration slider is moved), 3) pdf (students are reading the instructions from the ranking task), 4) concentration lab (students are interacting with the second simulation available directly from the Beer's law lab), 5) breaks, 6) other.}

\textbf{State-Action Sequences}. After the categorisation step, each $e_m \in (e_m, s_m, d_m)$ will have been replaced by $e_m'$, its corresponding event category, and each $s_m \in (e_m, s_m, d_m)$ will have been replaced by $s_m'$, its corresponding state mapping. For each timestep $m$, we encode $s'_m$ as a one-hot-encoded vector $v^{s'}_m$ of $c_s$ cells (with $c_s$ denoting the number of different state categories for the simulation), where all entries are $0$ except the one corresponding to $s'_m$, which is equal to $1$. We also transform $e'_m\in(e'_m, s'_m, d_m)$ into a one-hot-encoded vector $v^{e'}_m$ of $c_e$ cells (with $c_e$ denoting the number of different event categories for the simulation), where all entries are $0$ except the one corresponding to $e'_m$, which is equal to $d_m$. Finally, for each time step $m$, we replace each triplet $(e'_m, s'_m, d_m)$ by the concatenation of both vectors $v^{s'}_m$ and $v^{e'}_m$. 
For each student $u\in\mathcal{U}$, the interaction sequence $R_u$ can now be formulated as $\{[v^{s'}_0, v^{e'}_0], ..., [v^{s'}_{t_u-1}, v^{e'}_{t_u-1}]\}$.
We then normalise the sequences on the $v^{e'}_0$ dimension resulting in the final action sequence $F^{SA}_u$ for student $u\in\mathcal{U}$.

\vspace{0mm}
\subsection{Model Building}
In this paper, we are interested in creating a conceptual understanding prediction model that can accurately predict the conceptual understanding label $L_u$ for student $u$, given the extracted features $F_u^{SA}$. To this end, we rely on two types of models, both based on a \textit{Gated Recurrent Unit} (GRU) network \cite{cho2014properties}. GRUs are a type of Recurrent Neural Networks (RNNs), whose specificity is to allow the transmission of information over time. To process sequential data, RNNs are composed of $r$ recurrent cells arranged successively such that the output of one becomes the input of the next one, as illustrated on Figure~\ref{fig:grudl}. Their relative simplicity enables GRUs to handle smaller datasets (as those in our hands) than the more complex long short-term memory cells (LSTM), which contain an additional output gate~\cite{lstmchung2014empirical, lstmgruber2020gru}.

The first type of model is based on a GRU combined with a Dense layer (\texttt{GRU}), as shown in Figure \ref{fig:grudl}. Specifically, the extracted features $F^{SA}$ are fed into a neural architecture composed of a \emph{GRU}, a \emph{dropout} layer and a \emph{Dense} layer (with Softmax activation) having a hidden size of $1$. The model outputs the probability the student $u$ will gain advanced (label $1$) or limited (label $0$) understanding.

The second type of model is based on a Self-Attention-based Gated-Recurrent-Unit (\texttt{SA-GRU}), as shown in Figure \ref{fig:sagrudl}. In 2014, the concept of an \textit{Additive Attention} layer, also called \textit{Bahdanau attention} was introduced to improve statistical machine translation \cite{bahdanau2014neural}. Rather than using it to understand the context of words across languages, we use it to identify the pivotal moments in students' interactions with the simulation. We first take the raw extracted features $F^{SA}$ as the \textit{query}, \textit{key} and \textit{value} input matrices to our self-attention mechanism implemented as in \cite{attentionimplementation}. We then concatenate the attention output with the input features along each time step and feed this matrix to a GRU layer. Its latest output will directly be given to the dropout layer before passing through the final Dense layer which uses a Softmax activation (analogously to the \texttt{GRU} model). Though more complex than \texttt{GRU} and consequently more subject to overfitting on smaller datasets, \texttt{SA-GRU}'s attention weights can be used to understand what dimension of our features are important for the final prediction, and when that importance is emphasised over time. These aspects are fundamental to enable interpretability on such complex yet effective models. 

\section{Experimental Evaluation}
\label{sec:results}
We evaluated our (early) prediction models on two data sets collected using the ranking task activities on the PhET Beer's Law Lab and Capacitor Lab simulations (see Section \ref{section:contextdata}). We used the data set that we collected using the Beer's Law Lab activity as an \textit{evaluation data set} to assess to what degree student interaction data within the simulation is predictive for their obtained conceptual understanding (RQ1) and whether our models are able to predict students' conceptual understanding at the end of the task \textit{early} on (RQ2). We then used the data that was collected by previous work \cite{cock2021early} using the Capacitor Lab activity as a \textit{validation data set} to assess the transferability of our proposed models to a different population and simulation (RQ3). In the following, we first describe the experimental setup used for our analysis, before describing each experiment in detail.

\subsection{Experimental Setup}
\label{sec:experimentalsetup}
We compared the predictive performance of our models on the two data sets to baseline models used in prior work \cite{cock2021early}.

\subsubsection{Data Sets}
\label{subsec:datasets}
The evaluation and validation data sets were collected in two different classroom experiments (see Table \ref{tab:demographics}).

\renewcommand{\arraystretch}{1.1}
\begin{table}[t]
\resizebox{\columnwidth}{!}{%
\begin{tabular}{@{}lll@{}}
\toprule
& \textbf{Beer's Law Lab}  & \textbf{Capacitor Lab} \\                                    \midrule
\textbf{Number of students} & $254$ & $193$ \\
\midrule
\multirow{3}{*}{\textbf{Gender}} & Male: $53\%$ & \\ 
& Female: $43\%$ & N/A\\ 
& Other: $4\%$ &  \\                                                                        \midrule
\multirow{2}{*}{\textbf{Region}}  &A: $70\%$ & \multirow{2}{*}{USA}\\ 
& B: $30\%$ & \\                                                                            \midrule
\textbf{Education Level} & Vocational School & University \\                              \midrule
\multirow{2}{*}{\textbf{Field}} & Chemistry: $88\%$ & Physics majors: $17\%$\\
& Other: $12\%$ & Science/Engineering: $83\%$\\
\midrule
\textbf{Mean time in simulation} & 507s & 512s \\                                         \midrule
\textbf{Percentage label $1$}& $44\%$ & $51\%$\\
\bottomrule
\end{tabular}
}
\caption{Statistics of the data sets used for the experiments.}
\label{tab:demographics}
\end{table}
\renewcommand{\arraystretch}{1.0}

\textbf{Evaluation Data Set}. Using the Beer's Law Lab activity (see Section \ref{section:contextdata}), we collected data from $448$ laboratory technician apprentices of $10$ different vocational schools in a European country. The activity was performed directly in the classroom. We recorded students' interaction logs (clickstream data) as well as their responses to the ranking tasks. For our experiments, we removed participants who did not answer the ranking task ($23$ students), with an extremely short task duration (less than $2$ minutes), or with a low number of events (less than $10$ events). The final data set used for the experiment therefore consists of $254$ students. Prior to data collection, participants gave their informed consent to sharing the data for research, and all data was recorded in a completely anonymous way. Students could also participate in the learning activity without sharing their data for research. This study was approved by the responsible institutional review board (HREC number: 064-2021).

\textbf{Validation Data Set}. We used the data set collected by \cite{cock2021early} on the Capacitor Lab Activity (see Section \ref{section:contextdata}). It contains the interaction logs (clickstream data) and the ranking task responses of $193$ undergraduate Physics students of a US university. Students were all in their first year of studies and completed the activity as part of a mandatory homework assignment. Data was recorded in completely anonymous way and students had to give their informed consent prior to data collection. Moreover, students could participate in the learning activity without sharing their data for research. This study was approved by the responsible institutional review board (HREC number: 050-2020).

\subsubsection{Baseline Models}
\label{subsec:baseline}
We assessed the performance of our sequential models to more simple baseline models. Random forest (RF) models were successfully used in previous work \cite{cock2021early} to (early) predict students' level of conceptual understanding gained through interacting with an interactive simulation. \cite{cock2021early} experimented with neural network models and RF models with different type of features and did not find significant differences regarding predictive performance of the different model and feature combinations. We selected RF model with \textit{Action Span} features (\texttt{RF}) as a baseline for our experiments, as this encoding is most similar to our action-state sequences. The \textit{Action Span} features can be easily built from our state and action categories. For each time step $m$, the encoding of the triplet $(e'_m, s'_m, d_m)$ is similar to a one-hot encoded vector, where all entries are $0$ except the one representing $e'_m \times s'_m$ (state-action combination), which is equal to $d_m$. We therefore obtain a vector $v_m$ of dimensions $c_s \times c_e$. The raw interaction sequence $R_u$ for student $u\in\mathcal{U}$ then becomes $I_u = \{v_0, ..., v_{\tilde{n}_u\}}$. We flatten $I_u$ by summing all vectors $v_m$, resulting in a vector where each cell represents the amount of time in seconds the student spent doing each type of state-action combination. We then normalise the vector by the total duration of the interaction so that the model does not consider the length of each interaction, but rather the proportion of time spent on each condition.

\subsubsection{Optimisation Protocol}
Both \texttt{GRU} and \texttt{SA-GRU} are relatively sensitive to \textit{seeds} which have an influence on all weight initialisation. To temper the effects of randomness on the models' training, we used a cross validation with seed mitigation to optimise the neural network models. The usual nested cross-validation optimisation protocol was conducted for the baseline RF model. We evaluated performance of all models using the macro-averaged area under the ROC curve (AUC). 

\textbf{Optimisation for RF}. We optimised the \texttt{RF} model using a 10-fold nested cross validation stratified over the labels $L_u$. We use this step as an evaluation mean for the RF and as a basis for parameter optimisation for the GRU-based models. We ran the inner grid search on the following hyperparameters: the \textit{number of decision trees} $[3, 7, 9, 11, 13, 15, 17]$, the \textit{split criterion} $[gini, entropy]$, the \textit{maximum depth} $[5, 7, 9, 11, 13]$, the \textit{minimum samples split} $[3, 5, 7, 9, 11]$.

\textbf{Optimisation for GRU and SA-GRU}. For the two sequential models, we performed initial protoyping using a 10-fold nested cross validation (using the same folds as for the RF model) and a small number of seeds. We trained both the \texttt{GRU} and \texttt{SA-GRU} for $150$ epochs, using a categorical crossentropy loss and an Adam optimiser. We tuned the following hyperparameters: dropout rate after the GRU layer $[0.02, 0.05]$ and the number of cells $[32, 64]$ in the GRU layers. The initial experiments demonstrated that: the selected hyperparameter values in the inner grid search were the same for a majority of the folds, the categorical cross entropy loss on the inner validation data set tended to converge after $20-30$ iterations, and predictive performance of the models seemed to vary depending on the seed. We therefore decided to choose a fixed architecture for the \texttt{GRU} and \texttt{SA-GRU} models and limit the number of epochs to $30$ for all our experiments. For Beer's Law Lab the selected parameters were a dropout rate of $0.02$ and $32$ units for the GRU layer. For the Capacitor Lab, we also use $32$ units in the GRU Layer and a dropout rate of $0.05$. We evaluated the models using 10-fold cross validation (based on the same outer folds as the RF model). To mitigate the potential effects of randomness, we trained both models over $n$ different seeds and averaged predictive performance across seeds. For our experiments on the full state-action sequences (see Sections \ref{sec:resrq1} and  \ref{sec:resrq3}), we used $n=60$ seeds. Since we observed only minor differences in predictive performance across seeds, we reduced the number of seeds to $n=10$ for our early prediction experiments (see Sections \ref{sec:resrq2} and  \ref{sec:resrq3}).

\subsection{RQ1: Full Sequence Prediction}
\label{sec:resrq1}
In a first experiment, we investigated whether our proposed models were able to predict students' level of conceptual understanding based on their interaction data. We used the evaluation data set collected with the Beer's Law Lab simulation to answer RQ1.

\begin{figure}[!t]
  \centering
  \includesvg[width=0.9\columnwidth]{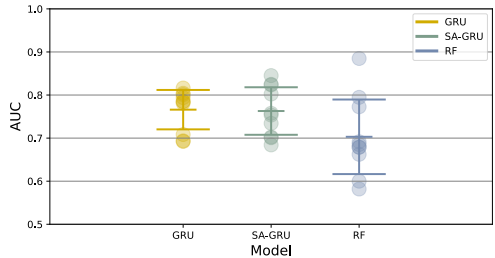}
  \caption{AUC of the \texttt{GRU}, \texttt{SA-GRU}, and \texttt{RF} models on the evaluation data set collected from Beer's Law Lab. Predictions were made based on the complete sequential interaction data of the students.}
  \label{fig:chemlabfullpred}
\end{figure}

\textbf{Predictive Performance}. Figure \ref{fig:chemlabfullpred} illustrates the predictive performance in terms of AUC for the \texttt{GRU}, the \texttt{SA-GRU}, and the \texttt{RF} baseline. Both sequential models outperform the \texttt{RF} baseline: using \texttt{GRU} and \texttt{SA-GRU} models we improve the AUC by $~10\%$ ($AUC_{GRU} = 0.77$, $AUC_{SA-GRU} = 0.76$, $AUC_{RF} = 0.70$). We also observe that the \texttt{RF} model exhibits a higher variability across folds than the two sequential models. For those, the minimum AUC is close to $0.70$ ($AUC_{min,GRU} = 0.69$, $AUC_{min,SA-GRU} = 0.68$). In contrast, performance of the \texttt{RF} model drops below $0.6$ based on the fold ($AUC_{min,RF} = 0.58$). We also performed a preliminary analysis on model bias according to protected attributes. Specifically, we grouped the students based on the available protected attributes and computed predictive performance in terms of AUC separately for the different groups. We found that the \texttt{GRU} model is slightly biased towards male students ($AUC_{m,GRU}=0.78$, $AUC_{f,GRU}=0.73$), while the two other models do not show any gender bias. All models show a bias in terms of region, with a higher AUC for region A ($AUC_{A,GRU}=0.78$, $AUC_{A,SA-GRU}=0.78$,$AUC_{A,RF}=0.76$) than for region B ($AUC_{B,GRU}=0.69$, $AUC_{B,SA-GRU}=0.66$,$AUC_{B,RF}=0.63$).

\begin{figure}[t]
  \centering
  \includegraphics[width=\columnwidth]{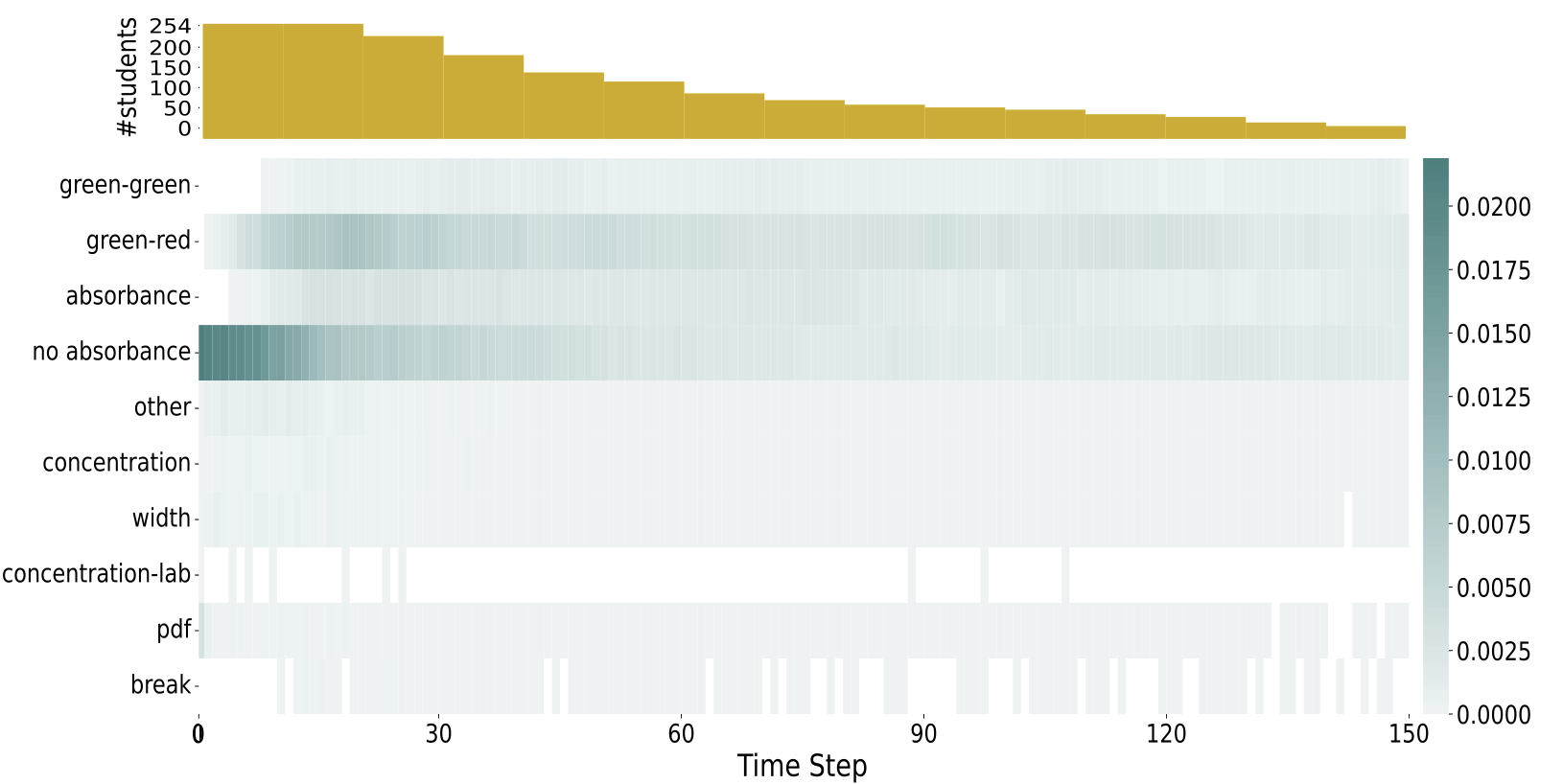}
  \caption{Normalized attention scores for the \texttt{SA-GRU} over $150$ time steps (bottom). Number of students with $\mathbf{n_u \geq t}$ for $\mathbf{t = 1,...,150}$ (top).}
  \label{fig:attheatmapfullchemlab}
\end{figure}

\textbf{Attention Interpretation}. While adding self-attention to the GRU does not increase predictive performance of the model (there is no significant difference in the performance of the \texttt{GRU} and \texttt{SA-GRU} model), the self-attention scores enable interpretation of the such black-box models. The heatmap in Fig. \ref{fig:attheatmapfullchemlab} shows the normalised attention scores for the \texttt{SA-GRU} model over $150$ time steps. For each student $u\in\mathcal{U}$, we extracted the output (dimension: $n_u \times f$) of the self-attention layer of the \texttt{SA-GRU} model (see Fig. \ref{fig:sagrudl}) and normalised it to make the attention scores comparable between models. We then averaged across all students. The number of students decreases rapidly with an increasing number of time steps (Fig. \ref{fig:chemlabfullpred} (top)). In fact, for $90\%$ of the students, $n_u \leq 150$.

We observe that the model seems to pay most attention to two of the states (\textit{green-red} and \textit{no absorbance}). The former state, \textit{green-red}, indicates a green laser colour and a red solution colour while the absorbance is measured. As expected, exploring with this parameter setting is important for solving the ranking task. The latter state, \textit{no absorbance}, indicates that the absorbance (outcome variable of the task) was not displayed, which makes solving the task impossible. Also the two other states (\textit{green-green} and \textit{absorbance}) seem to have phases of higher importance over time. For the \textit{no absorbance} state, we can see that the the model seems to pay less attention to it with an increasing number of time steps. Generally, the actual state of the simulation seems to be considered as more important than a specific action of the student. Regarding the action, the model seems to pay attention to state \textit{pdf}, which refers to students checking the task description and hence the configurations to be ranked. Also experimenting with \textit{concentration} and \textit{width} seems to have some importance. Interestingly, the importance of taking a \textit{break} varies over time steps. Finally, the model does not pay attention to the \textit{concentration-lab} action.

\textit{In summary, this first experiment demonstrated that our proposed models yielded a $10\%$ increase in AUC over the RF approach suggested in previous work \cite{cock2021early}. Furthermore, the sequential models also exhibit a lower variance across folds. Finally, by adding self-attention to a GRU model, we were able to provide interpretations of what type of student exploration behavior will lead to successfully solving the task.}

\subsection{RQ2: Early Prediction}
\label{sec:resrq2}
In our second experiments, we were interested in predicting students' level of understanding \textit{during} their interaction with the simulation. We therefore did not use all students' full interaction sequences, but used partial interactions up to a time step $l$ \jade{when the student went over that limit}. Therefore to build the features $F^{l,SA}_u$ for student $u\in\mathcal{U}$, we only used the interactions $I^l_u$ as a basis for building the features (see also Section \ref{sec:featextract}). We trained all our models for $l = 30, 40, 50, 60$. We did not make predictions earlier as we did not expect a model to be able to accurately predict conceptual understanding after only a few student actions. We further limited $l \leq 60$ as the number of students decreases for higher time steps (see also Fig. \ref{fig:attheatmapfullchemlab} (top) \jade{which indicates the amount of remaining students per time step}). At each time step $l$, we made predictions for students with state-actions sequences at each length $l$ (e.g. $\tilde{n}_u \geq l$). For students with shorter sequences (e.g., $\tilde{n}_u < l$), we used the last available prediction. We used the evaluation data set collected with Beer's Law Lab to answer RQ2.

\begin{figure}[t]
  \centering
  \includesvg[width=\columnwidth]{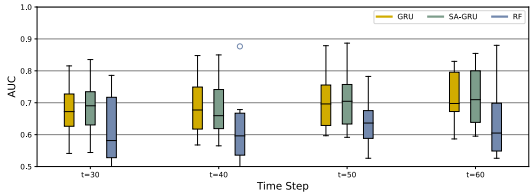}
  \caption{Early predictive performance in terms of AUC for \texttt{RF}, \texttt{GRU}, and \texttt{SA-GRU} for the evaluation data set collected using Beer's Law Lab for $\mathbf{t = 30, 40,...,60}$.}
  \label{fig:earlypredchemlab}
\end{figure}

\textbf{Predictive Performance}. Fig. \ref{fig:earlypredchemlab} illustrates the AUC of the different models for early prediction at $t = 30,40,50,60$. We again observe that the \texttt{RF} baseline model performs inferior to the sequential models. It achieves an AUC of $0.59$ after $30$ time steps and manages to only slightly improve over time with an AUC of $0.61$ after $60$ time steps. The AUC of the \texttt{RF} is $0.70$ for full sequence prediction (see Section \ref{sec:resrq1}), we therefore observe a strong increase in predictive performance when observing full sequences. It thus seems that the \texttt{RF} classifiers is not able to extract strong signal from early observations. The use of more elaborate sequential models leads to an up to $15\%$ increase in AUC ($AUC_{30,GRU} = 0.67$, $AUC_{30,SA-GRU} = 0.68$, $AUC_{60,GRU}$ $= 0.72$, $AUC_{60,SA-GRU} = 0.72$). Similar to the full sequences case, \texttt{RF} exhibits a much higher variance across folds than that our proposed models.

\textbf{Attention Interpretation}. Next, we investigated whether attention scores differed between an early prediction and a full sequence prediction model, i.e. whether the model paid attention to different features when making early predictions.

\begin{figure}[t]
  \centering
  \includegraphics[trim = 0cm 0.7cm 8.5cm 0cm, clip,width=\columnwidth]{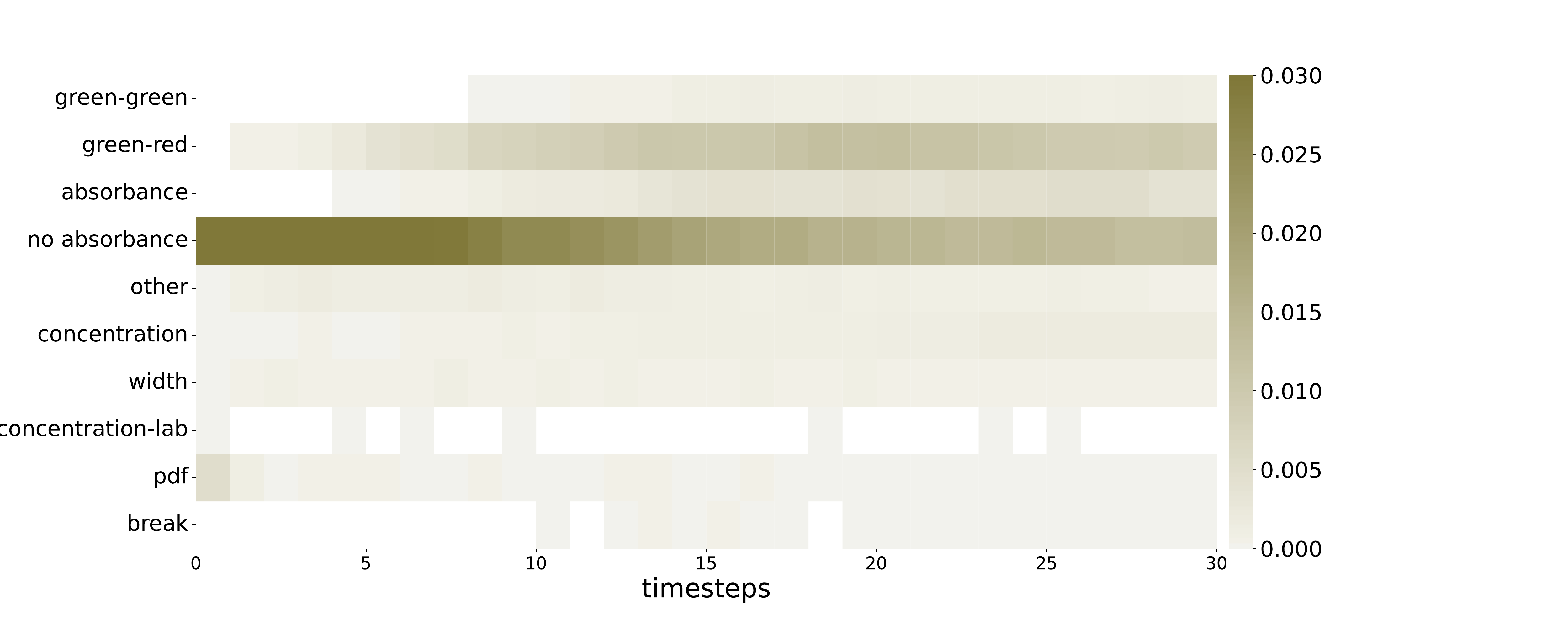}
  \caption{Normalised attention scores for \texttt{SA-GRU} trained for predicting at $\mathbf{t = 30}$ time steps for Beer's Law Lab.}
  \label{fig:attheatmap30ChemLab}
\end{figure}

Figure \ref{fig:attheatmap30ChemLab} shows the normalised attention scores for the \texttt{SA-GRU} model trained to predict at time step $t=30$ The scores in the heatmap were calculated following the same procedure as for the full sequences case (see Section \ref{sec:resrq1}). Similar to our findings on full sequences, we observe that the states generally seem to be assigned higher scores than the actions. While we again observe that the \textit{no absorbance} and \textit{green-red} states have relatively higher scores, we also observe that these scores vary over time. The scores of the \textit{no absorbance} state clearly decrease over time. This finding might suggest that many students realise during their first $10-15$ interactions with the simulation that the absorbance display should be turned on (by default, transmittance is displayed). In contrast, the scores of the \textit{green-red} state increase over time, indicating that after being able to observe the outcome variable (absorbance), it is important to select the optimal colours for laser (green) and solution (red). For the actions, we observe the same picture as for the full predictions. The \textit{concentration-lab} action gets little attention (it is probably also a rare action), while manipulating \textit{width} and \textit{concentration} gets similar attention over time. Checking the task description is mainly important early on. The model does not pay attention to \textit{break} for the first $10$ time steps. We hypothesise that ``thinking'' breaks do not occur often during initial interaction when students are still focused on understanding the task and have not yet started their exploration.\\

\textit{In summary, our second set of experiments showed that in case of early prediction, the use of a \texttt{GRU} or \texttt{SA-GRU} model leads to even larger performance increases than in case of full sequences. Depending on the time step, we observe improvements in AUC of $15\%$ compared to the RF baseline. Furthermore, the \texttt{RF} again shows high variability with drops in AUC down to $0.50$. We are again able to provide meaningful interpretations of the attention weights, which is a promising basis for targeted intervention.}

\subsection{\jade{RQ3: Generalisation}}
\label{sec:resrq3}
In our last experiment, we were interested in assessing the generalisability of the developed models. While both learning activities were based on a ranking task, the underlying simulation as well as the population are different. We therefore evaluated our models on the transfer data set collected from the Capacitor Lab simulation.

\begin{figure}[!t]
  \centering
  \includesvg[width=\columnwidth]{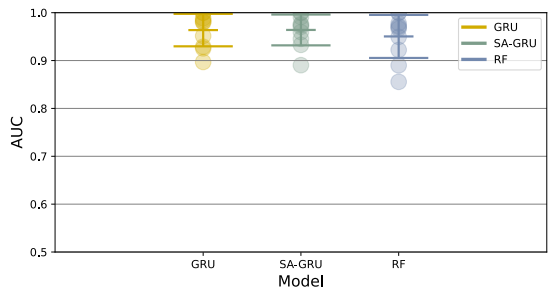}
  \caption{AUC of the \texttt{GRU}, \texttt{SA-GRU}, and \texttt{RF} models on the validation data set collected from CapacitorLab. Predictions were made based on the full interaction data of the students.}
  \label{fig:fullpredcap}
\end{figure}

\textbf{Prediction on Full Sequences}. We first evaluated the generalisability of the different model by predicting on the full interaction sequences of the validation data set. Figure \ref{fig:fullpredcap} illustrates the predictive performance in terms of AUC for the \texttt{GRU}, the \texttt{SA-GRU}, and the \texttt{RF} baseline. We observe that all the models achieve a very high AUC for the Capacitor Lab data set. The sequential models achieve an AUC of $0.96$ (\texttt{GRU}) and $0.96$ (\texttt{SA-GRU}) respectively, while the AUC of the \texttt{RF} baseline is $0.95$. Again the \texttt{RF} baseline model shows higher variability, achieving a minimum AUC of $0.86$. For the \texttt{GRU} and \texttt{SA-GRU}
models, we achieved a minimum AUC of $0.9$ and $0.89$ respectively. For Capacitor Lab, we ran the same bias analysis on protected attributes as for Beer's Law Lab, showing that models are not biased with respect to field (the only known protected attribute for this data set).
 
\begin{figure}[!t]
  \centering
  \includegraphics[width=\columnwidth]{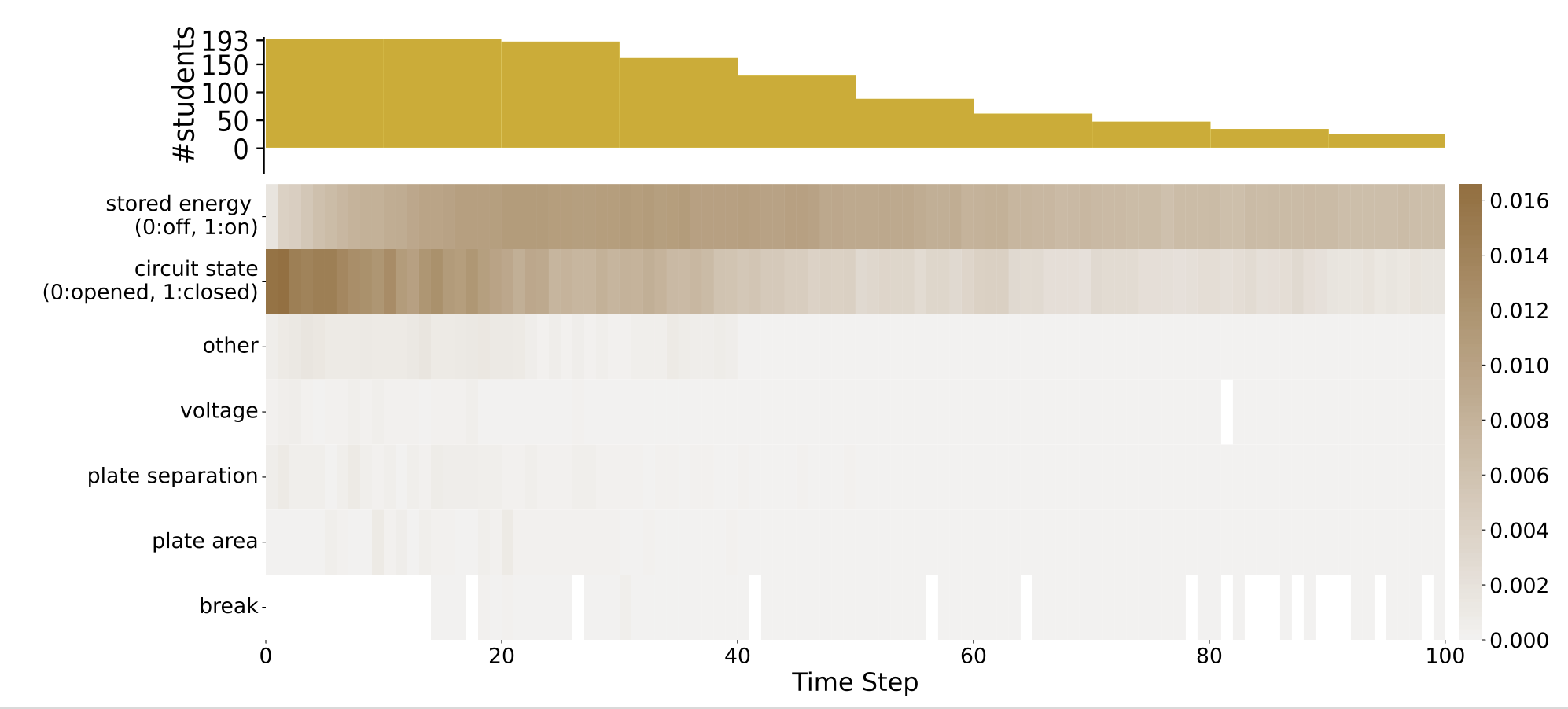}
  \caption{Normalised attention scores for CapacitorLab for the \texttt{SA-GRU} over $150$ time steps (bottom). Number of students with $\mathbf{n_u \geq t}$ for $\mathbf{t = 1,...,150}$ (top).}
  \label{fig:attheatmapfullcapacitor}
\end{figure}

While the \texttt{SA-GRU} model does not outperform the other two models, we again extracted the attention weights of this model and interpreted them. Figure \ref{fig:attheatmapfullcapacitor} shows the normalised attention scores of the \texttt{SA-GRU} models over a $100$ time steps. We chose to limit to $100$ time steps as for $90\%$ of the students $\tilde{n}_u \leq 100$. Again, the scores displayed in the heatmap were calculated using the procedure described in Section \ref{sec:resrq1}. Similar to our findings on Beer's Law Lab, we observe that the states (\textit{stored energy}, \textit{closed circuit}) receive higher attention than the actions. For \textit{stored energy}, attention scores increase up to time step $30$ and then decrease again after time step $60$. We assume that students needed some time to explore the simulation and figure out that the stored energy display needs to be turned on (it is by default turned off in the simulation). For the \textit{closed circuit} state, attention scores decrease over time, but there are always phases of increased activity. In order to be able to solve the ranking task, it is important to experiment with the closed as well as the open circuit, as detected by the model. The model also seems to pay attention to actions related to \textit{voltage}, \textit{plate area}, and \textit{plate separation}, which are the other three components in the simulation that influence the stored energy. Moreover, \textit{breaks} are important for different phases within the sequences. Hence, the model seems to detect when students take a thinking break and considers that as important.

\begin{figure}[!t]
  \centering
  \includesvg[width=\columnwidth]{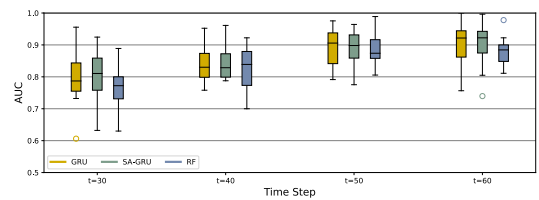}
  \caption{AUC of the \texttt{GRU}, \texttt{SA-GRU}, and \texttt{RF} models for early prediction ($\mathbf{t = 30, 40, ..., 60}$) on Capacitor Lab.}
  \label{fig:earlypredcaplab}
\end{figure}

\textbf{Early Prediction}. Next, we also trained the different models for early prediction on Capacitor Lab. Figure \ref{fig:earlypredcaplab} shows the AUC of the evaluated models for prediction on $t=30,40,50,60$ time steps. We see that the AUC of the two sequential model is close to $0.80$ already after observing $30$ student interactions ($AUC_{30,GRU} = 0.8$, $AUC_{30,SA-GRU} = 0.81$). Predictive performance of the \texttt{RF} model is slightly lower ($AUC_{30,RF} = 0.76$). After $60$ time steps, all the models achieve an AUC larger than $0.8$ ($AUC_{60,GRU} = 0.9$, $AUC_{60,SA-GRU} = 0.9$, $AUC_{60,RF} = 0.88$). We observe that all three models show a similar variability under this task.

\textit{In summary, our results demonstrate that the proposed models can be transferred to a different learning environment and population. In case of full interaction data, the performance of the models is inline with the \texttt{RF} baseline, in case of early prediction, the sequential models outperform the baseline. Furthermore, the addition of self-attention again enabled us to identify the key points in the exploration process.}
\section{Discussion and Conclusion}

Interactive simulations used for inquiry-based learning activities are increasingly gaining traction as part of science education at all levels. It has thus become essential to understand how students navigate in such environments in order to provide them with targeted and well-timed feedback. To address this issue, we introduced a new data-driven approach to predict students' conceptual understanding after solving inquiry-based learning tasks with interactive simulations. We leveraged log data directly recorded in these environments to extract meaningful features that were then used as input to models based on Gated Recurrent Units (\texttt{GRU}) to predict their conceptual understanding. In addition to a standard implementation of a GRU model, we also explored an extension, the Self-Attention-based Gated Recurrent Unit (\texttt{SA-GRU}) model, and compared their performance against a random forest (\texttt{RF}) classifier presented in prior work \cite{cock2021early}. We evaluated our approach on two data sets collected from different populations (vocational and undergraduate students) who used two different environments (a chemistry and a physics simulation). We aimed at answering to three research questions: 1) To what degree is student interaction data within interactive simulations predictive for their obtained conceptual understanding? 2) Can our models predict students' conceptual understanding early on? 3) Is our approach transferable to another context involving a different population and simulation?

To answer the first research question, we trained the proposed models using data collected from $254$ first, second and third year lab technician apprentices who solved a ranking task using the PhET Beer's Law Lab simulation. Using the complete log sequences of the students, both the \texttt{GRU} and \texttt{SA-GRU} models yielded AUC values of $0.77$ and $0.76$ respectively, outperforming the \texttt{RF} baseline model presented in previous work \cite{cock2021early} by $10\%$. For both models, we also observed lower variance across training folds, indicating a higher robustness compared to the \texttt{RF} baseline. Previous research has highlighted that along with predictive accuracy, it is the interpretability of the models that can make a significant difference in improving student learning outcomes \cite{conati2018ai}. While adding the self-attention layer to the GRU models did not further improve their predictive performance, our analyses illustrated that they significantly enhanced the interpretability of the model results. 
\jade{For instance, analysing the attention scores of the \texttt{SA-GRU} model, we observed that the state $s$ in which an action $a$ is conducted is more important than the nature of action $a$ itself to predict conceptual understanding. Experimenting in a setting where the variable of interest is observed, and where the environment settings correspond to those present in the problem, is influential.} 

Our findings are in line with those of \cite{wang2021automating}, who suggested that knowledge-grounded features that integrate qualitative observations are better predictors than purely actions-based features. From the attention score analysis, we also observed that taking breaks showed an interesting pattern with an alternating importance for the model over time. This results seems to indicate that taking breaks at specific moments during the activity may be predictive of the acquired understanding. These results are very much in line with those reported in previous work \cite{perez2017identifying,bumbacher2015learning}, who illustrated that recurrently taking breaks for reflection and planning can be a powerful inquiry strategy in interactive simulations.

To address the second research question, we then trained the models with only initial sequences of the students' log data in order to evaluate their capabilities to early predict students' conceptual understanding. Naturally, the performance of all models dropped compared to the training on the full sequences. However, both the \texttt{GRU} and \texttt{SA-GRU} outperformed the \texttt{RF} baseline also in early prediction, with AUC values that were up to $15\%$ higher and with lower variance across training folds. After only $30$ time steps in the simulation, AUC values for the \texttt{GRU} and \texttt{SA-GRU} models already reached $0.67$ and $0.68$ respectively. Moreover, the analysis of the attention scores of the \texttt{SA-GRU} model showed that already after $30$ time steps, meaningful interpretations could be made similar to those in the case of the models trained on the full sequences. This is a crucial result, since it enables the implementation of targeted interventions that could, for instance, support struggling students early on while they are still solving the tasks. Such interventions could potentially be implemented as automatic feedback that is directly displayed in the simulation, similar to approaches that have been adopted previously for other computer-based learning environments \cite{bimba2017adaptive}. Another possibility could be to display the insights gained from the models to teachers using real-time dashboards \cite{tavares2019towards,lopez2018dashboard},  to support them in preparing interventions for individual students and/or the entire class.   

Finally, to answer our third research question we evaluated our approach on a second data set from \cite{cock2021early}. It comprised log data from a different simulation (the PhET Capacitor Lab) that was used by a different population (undergraduate students attending a physics class). The goal of this analysis was to demonstrate that the functioning of our models is not restricted to a single context, but could be transferred to other populations and environments. In the case of training on the full sequences of log data, all three models performed very well with AUC values above $0.95$. Since in this case the \texttt{RF} baseline already showed a good performance, the use of the \texttt{GRU} and \texttt{SA-GRU} models only increased the AUC by $1\%$. Similar to the results from the Beer's Law Lab data set, we found higher importance for states (stored energy toggled and circuit closed) in the attention scores of the \texttt{SA-GRU} model. The high attention scores for the stored energy state could be explained by the observations made in previous work using the same simulation. \cite{lopez2018dashboard} found that despite the stored energy being the target outcome variable of their task, not all students toggled the function to display it. Similarly, the stored energy state could have served as a useful feature to detect unproductive behaviour in our task. Moreover, while the design of the ranking tasks for both simulations was similar, the use of different circuit states (closed or open) in the Capacitor Lab might be more discriminatory for student performance than the color states in Beer's Law Lab. This could explain why for the Capacitor Lab the RF baseline also performed very well. Finally, when looking at early prediction, we observed again that \texttt{GRU} and \texttt{SA-GRU} exhibited better performances compared to \texttt{RF} despite the differences being smaller than for Beer's Law Lab. Variances across folds, however, were similar across models.

Our findings represent an important step towards a better understanding and modelling student behaviour in \jade{ranking-based inquiry} learning activities within interactive simulations. \jade{Though the presented task is very specific, its key characteristics can be retained and transferred onto other interactive environments.}
Furthermore, the capabilities of the presented models to detect (un-)productive behaviour early on pave the way for more targeted interventions carried out by teachers or the system itself. 

Nevertheless, this study also comes with certain limitations. \jade{First, the chosen labels for the Beer's Law Lab are only able to separate \textit{struggling students} from \textit{more advanced students}, giving us an indication of whether students can conduct inquiry rather than whether they have understood Beer's Law.} Then, our preliminary bias analysis has indicated that some models show a slight gender bias (in case of the Beer's Law Lab simulation) and all models exhibit a geographic bias (also for the Beer's Law lab simulation). Therefore, further work on bias mitigation is needed. \jade{This includes understanding where the differences in predictions come from, as well as identifying any other confounding factors}. Furthermore, while the presented methods were applied to two different data sets involving different contexts to illustrate its transferability, it can not be guaranteed that they are unreservedly applicable to any other inquiry-based learning situation involving interactive simulations. More research including larger and more diverse samples as well as different educational levels, simulations, and learning tasks are needed to consolidate the generalisability of our methods. This is particularly important in order to mitigate the risks of algorithmic bias that could be introduced. Furthermore, it needs to be acknowledged that measuring students' conceptual knowledge by means of a single outcome variable (i.e., the final ranking submitted) might be too limiting. While the number of possible rankings could be considered sufficiently high to filter out the effect of random answers, other measuring tools (such as quizzes or interviews) could be included to complement the analyses. Finally, it should be emphasised that before any interventions based on the proposed methods are being introduced, it is imperative that the main stakeholders (i.e., teachers and students) are involved. Only in this way the potential educational impacts of such approaches can be assessed, implying both positive and negative consequences for the concerned groups.

\newpage
\balance
\bibliographystyle{abbrv}
\bibliography{sigproc}  
\balancecolumns
\end{document}